\documentstyle[12pt]{article}
\begin{document}

\begin{titlepage}
\begin{flushright}
BA-98-53\\
November 24, 1998 \\
\end{flushright}

\begin{center}
{\Large\bf  Atmospheric and Solar Neutrino  \\
~~Oscillations in $\nu$MSSM and Beyond
\footnote{Supported in part by  DOE under Grant No. DE-FG02-91ER40626
and by NATO, contract number CRG-970149.}
}
\end{center}
\vspace{0.5cm}
\begin{center}
{\large Qaisar Shafi$^{a}$\footnote {E-mail address:
shafi@bartol.udel.edu} {}~and
{}~Zurab Tavartkiladze$^{b}$\footnote {E-mail address:
tavzur@axpfe1.fe.infn.it} }
\vspace{0.5cm}

$^a${\em Bartol Research Institute, University of Delaware,
Newark, DE 19716, USA \\

$^b$ Institute of Physics, Georgian Academy of Sciences,
380077 Tbilisi, Georgia}\\
\end{center}

\vspace{1.0cm}

\begin{abstract}

We show how a unified description of the various two-flavor neutrino
oscillation solutions, allowed by the atmospheric and solar neutrino
experiments, are naturally realized within the 
framework of $\nu$MSSM (MSSM augmented with the seesaw mechanism) and 
beyond, especially grand unified theories. 
A general mechanism for achieving maximal mixing to resolve the
atmospheric anomaly is discussed, and applied to the flipped $SU(5)$
model. 
Except in the case of MSSM and $SU(5)$, a light sterile neutrino is an
inevitable consequence of our considerations. 
The
bi-maximal neutrino mixing scenario is one of the options considered.
Neutrino hot dark matter can arise in models with
maximal $\nu_{\mu }-\nu_s$ oscillations. 
A ${\cal U}(1)$ flavor symmetry, motivated 
by the charged fermion mass hierarchies and the magnitudes of the CKM 
matrix elements, plays a central role. 

\end{abstract}

\end{titlepage}



The recent atmospheric neutrino results from 
Superkamiokande (SK) 
provide new, and increasingly credible evidence for the
existence of neutrino oscillations. The atmospheric neutrino data
seems consistent with a two-flavor model involving
$\nu_{\mu }-\nu_x$ oscillations, where $\nu_x$ could either be
$\nu_{\tau }$ or a new sterile state $\nu_s$. The data favors
a large ($\sin^2 2\theta \stackrel{_>}{_\sim } 0.8$) mixing angle, with
$\Delta m^2 \sim 10^{-2}-10^{-3}~{\rm eV}^2$ \cite{sup}. 
The solar neutrino data
is consistent \cite{bah} both with the vacuum solution in which
$\Delta m^2 \sim 5\cdot 10^{-11}-10^{-10}~{\rm eV}^2$ and
$\sin^2 2\theta \sim 0.7- 1.0$,
or the small mixing angle MSW solution with
$\sin^2 2\theta \sim $(few)$\cdot 10^{-3}$
and $\Delta m^2 \sim $(few)$\cdot 10^{-6}~{\rm eV}^2$.
The `vacuum' solution strongly indicates oscillations
between active
neutrinos, while the MSW solution can involve either an active or 
sterile neutrino \cite{bah}.

The importance of the atmospheric and solar neutrino experiments stems
from the fact that they may provide the first unequivocal evidence for
physics beyond the standard model. It is hard to see how neutrino masses
of order $10^{-1}$~eV (or greater) can be generated within the framework
of the standard model (SM) or its minimal supersymmetric extension (MSSM),
even after non-renormalizable interactions are taken into account.
Some extension of this framework is mandated, and perhaps the most
elegant possibility is to introduce right handed neutrinos which,
when coupled with the seesaw mechanism \cite{seesaw}, gives rise to neutrino
masses and mixings. 
There are well known extensions of the SM, such as
$SU(4)_c\times SU(2)_L\times SU(2)_R$ \cite{pati}, $SO(10)$ \cite{so10}
or flipped $SU(5)$ \cite{flip}, 
which all predict the existence of right-handed neutrinos, and consequently
non-zero neutrino masses and mixings.

A consistent theoretical framework for understanding the neutrino
oscillation parameters allowed by experiments should also shed light 
on the observed charged
fermion mass hierarchies and the magnitude of the CKM matrix elements.
It seems natural to think that there exists some `flavor' (horizontal)
symmetry that distinguishes between the generations.
In the supersymmetric setting, a global ${\cal U}(1)$-${\cal R}$-symmetry 
would provide an example
of such a symmetry. Note that the flavor symmetry need not be continuous, 
and indeed some superstring inspired models suggest the existence
of discrete flavor `gauge' symmetries.

In this letter we would like to develop a unified approach,
based on $\nu$MSSM (MSSM augmented with the seesaw mechanism) and its grand
unified extensions, that incorporates the various two-flavor
oscillation solutions allowed by the data.
Our approach is strongly guided by the desire to provide a general
mechanism for realizing (essentially) maximal mixing in 
$\nu_{\mu }-\nu_x$ oscillations, which seems to be favored experimentally,
to explain the atmospheric neutrino
anomaly. The scheme that we propose here exploits a ${\cal U}(1)$ 
flavor symmetry \cite{u1}, whose
breaking by a $\nu$MSSM singlet field $X$
yields an important `expansion' parameter
$\epsilon \equiv \langle X \rangle / M_P=0.2$, where
$M_P = 2.4\cdot 10^{18}$~GeV denotes the reduced Planck scale. 
In $\nu$MSSM the elements of the quark and lepton Yukawa matrices 
are expressed in powers of the expansion parameter $\epsilon$,
while in GUTs another dimensionless parameter 
$\epsilon_G(\equiv M_{GUT}/M_P)\sim 10^{-2}$ can also play a role.


A brief summary of our paper is as follows. Within the framework of
$\nu$MSSM supplemented by a ${\cal U}(1)$ flavor symmetry, 
we find (see also \cite{ram}) that the SK data can be
accommodated by implementing $\nu_{\mu }-\nu_{\tau }$ oscillations with
large mixing to explain the atmospheric anomaly, while the solar neutrino 
puzzle is resolved via the small mixing angle MSW solution. 
This scheme has the distinctive features that it does not require the
existence of any sterile neutrino, has no `hot' dark matter, and can
be extended, more or less straightforwardly, to $SU(5)$. One of its
disadvantages is that some tuning  of the
parameters (with $\sim 1\%$ accuracy)
is required in 
order to achieve the correct magnitudes for the mass squared differences
$\Delta m^2_{atm}$ and $\Delta m^2_{solar}$.
Perhaps a more serious shortcoming of this scheme is that it is
not clear how to extend it to other interesting gauge groups such as   
$SU(4)_c\times SU(2)_L\times SU(2)_R$, $SO(10)$ \cite{babu} 
and flipped $SU(5)$.

We propose a rather general mechanism for realizing
(essentially) maximal $\nu_{\mu }-\nu_x$ mixings to explain the 
atmospheric anomaly. This mechanism inevitably requires the existence
of a `light' sterile neutrino $\nu_s$ in order to provide a simultaneous 
resolution of the atmospheric and solar neutrino anomalies.
We show how the various two-flavor
neutrino oscillation scenarios are naturally realized within 
the $\nu$MSSM (more precisely perhaps $\nu_s$MSSM
to emphasize the presence of a sterile neutrino) framework.  
The existence of neutrino `hot' dark matter is noted  in models
that exhibit either bi-maximal neutrino mixing or maximal 
$\nu_{\mu }-\nu_s$ and small angle $\nu_e-\nu_{\tau }$ oscillations. 
The scenario here with
bi-maximal neutrino mixings differs from those considered in 
ref. \cite{bimax}, since it involves a sterile neutrino state. 
In order to demonstrate the applicability of our approach
to models beyond $\nu$MSSM, we show how it resolves the solar and
atmospheric neutrino puzzles in flipped $SU(5)$.



We begin with the quark sector and choose the ${\cal U}(1)$ flavor 
charge of the superfield $X$ to be $Q_X=1$. Expressed in terms 
of $\epsilon$, the magnitudes of the
CKM matrix elements $V_{us}$ and $V_{cb}$ are as follows:

\begin{equation}
V_{us}\sim \epsilon~,~~~
V_{cb}\sim \epsilon^2~,
\label{ckm}
\end{equation}
from which,
\begin{equation}
Q_{q_3}-Q_{q_2}=2~,~~~~~~~Q_{q_2}-Q_{q_1}=1~.
\label{Qq}
\end{equation}
Note that (\ref{Qq}) automatically gives $Q_{q_3}-Q_{q_1}=3$,
so that 
$V_{ub}\sim \epsilon^3$, which indeed is of the right magnitude.
From the known mass hierarchies of the up and down
quarks, we have the following ratios of the asymptotic Yukawa 
couplings:

$$
\lambda_u : \lambda_c :  \lambda_t \sim
\epsilon^6 : \epsilon^3 :1~,
$$
\begin{equation}
\lambda_d :\lambda_s :\lambda_b \sim 
\epsilon^4:\epsilon^2 :1~,
\label{lambdaud}
\end{equation}
which, taking into account (\ref{Qq}), gives

$$
Q_{u^c_3}-Q_{u^c_2}=1~,~~~~~~Q_{u^c_2}-Q_{u^c_1}=2~,
$$
\begin{equation}
Q_{d^c_3}=Q_{d^c_2}~,~~~~~~~~~~~~Q_{d^c_2}-Q_{d^c_1}=1~.
\label{Qqr}
\end{equation}

Expressions (\ref{Qq}) and (\ref{Qqr}) determine the structures
of the Yukawa matrices of the up and down quarks (throughout the paper
the matrix elements are only determined up to factors of order unity;
we also ignore CP violation):

\begin{equation}
\begin{array}{ccc}
 & {\begin{array}{ccc}
~~u^c_1& \,\,~u^c_2  & \,\,~u^c_3~~~
\end{array}}\\ \vspace{2mm}
Y_u \sim 
\begin{array}{c}
q_1\\ q_2 \\q_3 
 \end{array}\!\!\!\!\! &{\left(\begin{array}{ccc}
\,\,\epsilon^6  &\,\,~~\epsilon^4 &
\,\,~~\epsilon^3 
\\ 
\,\,\epsilon^5   &\,\,~~\epsilon^3  &
\,\,~~\epsilon^2
 \\
\,\, \epsilon^3 &\,\,~~ \epsilon  &\,\,~~1 
\end{array}\right) }~,
\end{array}  \!\!  ~~~~~
\begin{array}{ccc}
 & {\begin{array}{ccc}
~d^c_1& \,\,~d^c_2  & \,\,~d^c_3~~
\end{array}}\\ \vspace{2mm}
Y_d \sim
\begin{array}{c}
q_1\\ q_2 \\q_3 
 \end{array}\!\!\!\!\! &{\left(\begin{array}{ccc}
\,\,\epsilon^4  &\,\,~~\epsilon^3 &
\,\,~~\epsilon^3 
\\ 
\,\,\epsilon^3   &\,\,~~\epsilon^2  &
\,\,~~\epsilon^2
 \\
\,\, \epsilon &\,\,~~ 1  &\,\,~~1 
\end{array}\right) }~. 
\end{array}  \!\!  ~~~~~
\label{downup}
\end{equation}
As emphasized by several authors \cite{u1}, the flavor ${\cal U}(1)$ symmetry 
can provide a nice understanding of the quark mass hierarchies 
as well as the CKM mixing angles.  


Turning to the lepton sector, let us remark that
neutrino oscillations arise from a nontrivial lepton mixing matrix

\begin{equation}
V_{{\nu}CKM}=U_e^{\dag }U_{\nu }V_{\nu }~,
\label{nuckm}
\end{equation}
where
$U_e$ and $U_{\nu }$ respectively diagonalize the left-handed 
charged lepton and neutrino
`Dirac' mass matrices, and $V_{\nu }$ diagonalizes the matrix 
$m_D^{diag}M_R^{-1}m_D^{diag}$ ($m_D$ and $M_R$
denote the `Dirac' and `Majorana' mass matrices respectively).
Clearly, rotations of the left-handed states
$l_i=\left(e~,\nu \right)_i$ only partially determine the $V_{{\nu}CKM}$
mixing angles. It seems reasonable to begin the study of lepton
mixings with the charged lepton sector, since our knowledge
about masses in this sector is so well established. 

The well-known hierarchies among the Yukawa couplings 
of the charged leptons
can be written:
\begin{equation}
\lambda_e :\lambda_{\mu } :\lambda_{\tau } \sim
\epsilon^5:\epsilon^2 :1~,
\label{lamdae}
\end{equation}
from which we conclude that:

\begin{equation}
Q(l_3e^c_3)-Q(l_1e^c_1)=5~,~~~~~
Q(l_3e^c_3)-Q(l_2e^c_2)=2~.
\label{Qs}
\end{equation}
From (\ref{Qs}), through proper prescription of the charges of
$l$ and $e^c$
states, we obtain the following possible choices for the structure of
$Y_e$:

\begin{equation}
\begin{array}{ccc}
 & {\begin{array}{ccc}
\hspace{-5mm}~~e^c_1~ & \,\,~~~~e^c_2  & \,\,~~~e^c_3

\end{array}}\\ \vspace{2mm}
{Y}_e\sim \begin{array}{c}
l_1\\ l_2 \\l_3
 \end{array}\!\!\!\!\! &{\left(\begin{array}{ccc}
\,\,\epsilon^5~~~~  &\,\,\epsilon^{2+k} &
\,\,\epsilon^{n+k}
\\
\,\,\epsilon^{5-k}~~   &\,\,\epsilon^2~~  &
\,\,\epsilon^n~~
 \\
\,\,\epsilon^{5-n-k} &\,\,\epsilon^{2-n} &\,\,1~~
\end{array}\right) }~,
\end{array}  \!\!  ~~~~~
\label{lep3}
\end{equation}
where
\begin{equation}
0\leq n \leq 2~,~~~~~~~~0\leq k \leq 5-n~,
\label{cond}
\end{equation}
and the ${\cal U}(1)$ charges are
$$
Q_{l_3}=0~,~~~Q_{l_2}=-n~,~~~Q_{l_1}=-n-k~,
$$
\begin{equation}
Q_{e^c_3}=0~,~~~Q_{e^c_2}=n-2~,~~~
Q_{e^c_1}=n+k-5~.
\label{Qlep}
\end{equation}

For generating neutrino masses through seesaw mechanism one has to
introduce right handed $\nu^c$ states. In models such as
$SU(4)_c\times SU(2)_L\times SU(2)_R$, $SO(10)$ and flipped $SU(5)$
these states are embedded in the unified multiplets. Considering three
$\nu^c$ states with the following ${\cal U}(1)$ charges:

\begin{equation}
Q_{\nu^c_3}=0~,~~~Q_{\nu^c_2}=-n'~,~~~
Q_{\nu^c_1}=-n'-k'~,
\label{Qnur}
\end{equation}
the `Dirac' and `Majorana' mass matrices will have the forms:

\begin{equation}
\begin{array}{ccc}
 & {\begin{array}{ccc}
~~\nu^c_1& \,\,~~~~~~~~~~\nu^c_2  & \,\,~~~~~~~\nu^c_3~~~
\end{array}}\\ \vspace{2mm}
m_D=
\begin{array}{c}
\nu_e\\ \nu_{\mu } \\ \nu_{\tau }  
 \end{array}\!\!\!\!\! &{\left(\begin{array}{ccc}
\,\,\epsilon^{n+n'+k+k'}  &\,\,~~\epsilon^{n+k+n'} &
\,\,~~\epsilon^{n+k}
\\ 
\,\,\epsilon^{n+n'+k'}   &\,\,~~\epsilon^{n+n'} &
\,\,~~\epsilon^n
 \\
\,\, \epsilon^{n'+k'} &\,\,~~\epsilon^{n'}  &\,\,~~1 
\end{array}\right)h_u }~,
\end{array}  \!\!  ~~~
\label{dir}
\end{equation}

\begin{equation}
\begin{array}{ccc}
 & {\begin{array}{ccc}
\nu^c_1& \,\,~~~~~~~\nu^c_2  & \,\,~~~~~~~\nu^c_3~~
\end{array}}\\ \vspace{2mm}
M_R=
\begin{array}{c}
\nu^c_1\\ \nu^c_2 \\ \nu^c_3 
 \end{array}\!\!\!\!\! &{\left(\begin{array}{ccc}
\,\,\epsilon^{2(n'+k')}  &\,\,~~\epsilon^{2n'+k'} &
\,\,~~\epsilon^{n'+k'} 
\\ 
\,\,\epsilon^{2n'+k'}   &\,\,~~\epsilon^{2n'}  &
\,\,~~\epsilon^{n'}
 \\
\,\, \epsilon^{n'+k'} &\,\,~~\epsilon^{n'}  &\,\,~~1 
\end{array}\right)M }~, 
\end{array}  \!\!  ~~~
\label{maj}
\end{equation}
where $M$ is some mass scale.

From (\ref{dir}), the contributions
from $m_D$ to $V_{\nu CKM}$ (namely the elements of $U_{\nu }$) are
$\theta_{12}^{\nu }\sim \epsilon^k$,
$\theta_{23}^{\nu }\sim \epsilon^n$,
$\theta_{13}^{\nu }\sim \epsilon^{n+k}$. Note that the diagonalization
of the matrix $m_D$ does not change the hierarchical structure of the matrix $M_R$.
As far as the elements of the $V_{\nu }$ matrix are
concerned, they are determined from the matrix

\begin{equation}
\begin{array}{ccc}
 & {\begin{array}{ccc}
\nu_e& \,\,~~~~~~~\nu_{\mu }  & \,\,~~~~~~\nu_{\tau }~~
\end{array}}\\ \vspace{2mm}
m_{\nu }=m_D^{diag}M_R^{-1}m_D^{diag}=
\begin{array}{c}
\nu_e\\ \nu_{\mu } \\ \nu_{\tau } 
 \end{array}\!\!\!\!\! &{\left(\begin{array}{ccc}
\,\,\epsilon^{2(n+k)}  &\,\,~~\epsilon^{2n+k} &
\,\,~~\epsilon^{n+k} 
\\ 
\,\,\epsilon^{2n+k}   &\,\,~~\epsilon^{2n}  &
\,\,~~\epsilon^{n}
 \\
\,\, \epsilon^{n+k} &\,\,~~\epsilon^{n}  &\,\,~~1 
\end{array}\right)\frac{h_u^2}{M}}~, 
\end{array}  \!\!  ~~~
\label{mactive}
\end{equation}
which gives
$\theta_{12}'\sim \epsilon^k$,
$\theta_{23}'\sim \epsilon^n$,
$\theta_{13}'\sim \epsilon^{n+k}$. 
From (\ref{mactive})we have for the masses of the light neutrinos:

\begin{equation}
m_{\nu_3}\sim
\frac{h_u^2}{M}~,
m_{\nu_1}:
m_{\nu_2}:
m_{\nu_3}\sim
\epsilon^{2(n+k)}:
\epsilon^{2n}:
1~.
\label{nu132s1}
\end{equation}
The mixing angles and masses of the left handed neutrinos do not
depend on the ${\cal U}(1)$ charges of the $\nu^c$ states.

The case with $n=0$ leads to the desirable large 
$\nu_{\mu }-\nu_{\tau }$ mixing. According to (\ref{nu132s1}) this
case also gives 
$m_{\nu_{2}}\sim m_{\nu_{3}}$. To solve solar
neutrino puzzle through $\nu_e-\nu_{\mu }$ oscillations, we need some
fine tunings in order to realize a suitably small mass for the 
$`\nu'_{2}$ state. For $k=0$ the solar neutrino puzzle can be resolved
through the large angle vacuum oscillations, while $k=2$ would 
suggest small angle MSW oscillations. 
 However, in these cases fine
tunings to accuracies $\sim 0.01\%$ and $1\%$ respectively
must be done. To avoid this `unpleasant' fact, one can either introduce
a sterile neutrino state for solution of the solar neutrino puzzle
through small angle MSW oscillations, or perhaps put forward another mechanism
for keeping the $`\nu_2'$ neutrino light.
The two cases (with $(n,~k)=(0,~2)$ and $(n,~k)=(0,~0)$), as pointed out 
in \cite{ram}, can be realized in MSSM and $SU(5)$ GUT respectively.


In many realistic unified theories, however, the choice $n=0$ is not 
realized and so the contributions to the lepton mixing matrix
in (\ref{nuckm}) from the neutral sector are critical if the 
atmospheric neutrino anomaly is to be resolved. 
The neutrino `Dirac' mass matrices in models such as
flipped $SU(5)$,  $SU(4)_c\times SU(2)_L \times SU(2)_R$ and $SO(10)$
mostly have the same `hierarchical' structure as the 
mass matrices of up quarks
(or leptons in some special cases). This means that the contributions
from $V_{\nu }$ to the lepton mixing matrix will also be small 
if $n\neq 0$  
\footnote{The case $(n,~k)=(1,~1)$ in 
(\ref{lep3}) can arise from
$SU(4)_c\times SU(2)_L\times SU(2)_R$ and
flipped $SU(6)$ theories with realistic pattern of fermion masses
and mixings (see \cite{422} and \cite{61} respectively);
the case $(n,~k)=(2,~2)$ arises in a simple
version of flipped $SU(5)$ model which we consider below.}, 
unless the neutral sector of the models are suitably modified.  
As we will see, with our approach this also opens up 
the possibility of realizing the other allowed two-flavor 
oscillation solutions within the framework of realistic 
SUSY GUTs (as well as $\nu$MSSM).  The mechanism that we propose 
naturally leads to maximal mixing between two oscillating neutrino 
flavors, and turns out to be a powerful tool for model building. 
By this mechanism a quasi degenerate texture for the mass matrix of the
two flavors are obtained, and so the masses of the appropriate neutrino
states cannot be changed through fine tunings. For an explanation
of atmospheric and solar neutrino anomalies we led to introduce
a sterile neutrino state $\nu_s$.

The introduction of the sterile state opens up the following
scenarios (that are consistent with the latest solar and atmospheric
neutrino data):

$a)$ Atmospheric anomaly is explained through maximal 
$\nu_{\mu }-\nu_{\tau }$ mixing, while the solar neutrino puzzle is resolved 
by small angle MSW  $\nu_e-\nu_s$ oscillations.

$b)$ Deficit of atmospheric neutrinos is due to maximal  
$\nu_{\mu }-\nu_s$ mixings and maximal $\nu_e-\nu_{\tau }$
mixing resolves the solar neutrino anomaly (bi-maximal mixing scenario).

$c)$ Atmospheric neutrino anomalies still explained by maximal angle   
$\nu_{\mu }-\nu_s$ oscillations, while the solar neutrino puzzle 
is resolved through small angle $\nu_e-\nu_{\tau }$ MSW oscillations.
Below we will present a detailed discussion of how these scenarios are realized, and
their phenomenological implications in a model independent way. 


Let us demonstrate the mechanism by obtaining maximal  
$\nu_{\mu}-\nu_{\tau }$ mixing, but it can work for any two 
flavors\footnote{Especially interesting may be the case of maximal 
$\nu_e-\nu_{\tau }$
mixing which provides the vacuum oscillation solution for
the solar neutrino puzzle.}. We will show mechanism how to
obtain large mixing for $n\neq 0$ (this case can include the choice
$k=0$ ). We introduce two MSSM singlet superfields 
${\cal N}_{2,3}$,
which couple with $\nu_{e,\mu,\tau}$ to form the Dirac matrix.
Their ${\cal U}(1)$ charges are as follows:

\begin{equation}
Q_{{\cal N}_2}=-(p+r)~,~~~~~ Q_{{\cal N}_3}=p,
\label{QQs}
\end{equation}
where $p$, $r$ are integers, and
$0< p \leq n\neq 0$, $r\geq 0$ (it is important to have $Q_{{\cal N}_3}>0$
and $Q_{{\cal N}_3}+Q_{l_3}+Q_{h_u}>0$; we also took
$Q_{l_3}=Q_{h_u}=0$ for a simpler presentation).
The `Dirac' and `Majorana' mass matrices turn out to be
%
%

\begin{equation}
\begin{array}{ccc}
{m_D= \left(\begin{array}{ccc}
\,\,\epsilon^{n+k+p+r} &\,\,\epsilon^{n+k-p}
\\
\,\, \epsilon^{n+p+r} &
\,\,\epsilon^{n-p}
\\
\,\, \epsilon^{p+r}~~~ &\,\,0~~
\end{array}\right)\kappa \langle h_u\rangle }
\end{array}  \!\!~,~~~~~~~~
\begin{array}{cc}
{M_{{\cal N}}= \left(\begin{array}{ccc}
\,\, \epsilon^{2p+r}
 &\,\,1
\\
\,\, 1~~&\,\,0
\end{array}\right)M\epsilon^{r}}
\end{array}~.~
\label{mats}
\end{equation}
%
%
%
Here $\kappa $ and $M$ respectively denote some coupling constant
and mass scale and depend on the details of the model.
The light chiral neutrinos will acquire mass through the seesaw 
mechanism. Namely,

\begin{equation}
\begin{array}{cc}

m_{\nu }=m_DM_{{\cal N}}^{-1}
m_D^T=
\!\!\!\!\! &{\left(\begin{array}{ccc}
\,\,\epsilon^{n+2k} &\,\,\epsilon^{n+k}  &\,\,~~\epsilon^k
\\
\,\,\epsilon^{n+k} &\,\,\epsilon^n  &\,\,~~1
\\
\,\,\epsilon^k&\,\, 1 &\,\,~~0
\end{array}\right)m}~,~~~
\end{array}  \!\!  ~~~~~
m=\frac{\kappa^2\langle h_u\rangle^2}{M}\epsilon^n~.
\label{nu2}
\end{equation}
From (\ref{nu2}) we see that two states are quasi-degenerate with masses
$m_{\nu_2}\simeq m_{\nu_3}\simeq m$ and the third $\nu_1$ state is 
massless since the matrix (\ref{nu2}) is obtained by the exchange of the
two heavy singlet states. 
The matrix in (\ref{nu2}) generate  desired
(essentially) maximal mixing \cite{bin} in the 
$\nu_{\mu }-\nu_{\tau }$ sector. 

As may be expected, there is no dependence in (\ref{nu2}) on the
${\cal U}(1)$-charges of ${\cal N}_2$, ${\cal N}_3$.
For the neutrino oscillation parameters we find:

$$
\Delta m_{\mu \tau }^2=2m^2\epsilon^n,
$$
\begin{equation}
\sin^2 2\theta_{\mu \tau } = 1-{\cal O}(\epsilon^{2n})~,
\label{atm}
\end{equation}
For $n=1$, $m\sim 0.1$~eV, or $n=2$, $m\sim 0.2$~eV,
one can obtain a value for
$\Delta m_{\mu \tau }^2$($\sim 3\times 10^{-3}~{\rm eV}^2$) 
that is consistent with the atmospheric neutrino results.
For either case $n=1$ or $2$, the model must be arranged such that 
these values for $m$ are realized. Note that the neutrino masses
in both cases are much too small for any significant contribution 
to hot dark matter.

However, the masses are sufficiently large (!) that the solution 
of the solar neutrino
puzzle requires a sterile neutrino $\nu_s$ \cite{foot}.
The relevant superpotential couplings are:

\begin{equation}
W_{\nu_e s}=
\Gamma l_1\nu_sh_u+
m_{\nu_s}\nu_s^2~,
\label{wsol}
\end{equation}
The coefficients $\Gamma $ and $m_s$ should be strongly 
suppressed in order to guarantee a
light $\nu_e-{\nu_s}$ system. This can be realized  with
a $U(1)$-${\cal R}$ symmetry 
\cite{chun, 61, 422}.
With the oscillation parameters  
$$
\Delta m_{es}^2\simeq m_{\nu_s}^2
\sim 10^{-6}~{\rm eV}^2~,
$$
\begin{equation}
\sin^2 2\theta_{es}\simeq 4\left(\frac{\Gamma h_u}{m_{\nu_s}}\right)^2
\sim 6\cdot 10^{-3} ~.
\label{sol}
\end{equation}
one finds
\begin{equation}
m_{\nu_e \nu_s}\equiv \Gamma h_u\sim
4\cdot 10^{-5}~{\rm eV}~,~~
m_{\nu_s }\sim  10^{-3}~{\rm eV}.
\label{nust}
\end{equation}


The couplings of more heavy $\nu_{\mu }$ and $\nu_{\tau }$ states with
$\nu_s$ are expected to be of the order of 
$\frac{\Gamma }{\epsilon^{n+k}}(\nu_{\tau }+\epsilon^n\nu_{\mu })h_u$.
Taking into account (\ref{cond}) and (\ref{nust}) we will have
$\Gamma /\epsilon^{n+k}h_u\stackrel{_<}{_\sim }0.1$~eV, which do not
exceed
to $m$ and therefore the results presented above will be unchanged. 


The above discussion shows how one can obtain 
maximal mixing between two
`active' neutrinos, in this case $\nu_{\mu }$, $\nu_{\tau }$, 
which is in good 
agreement with the atmospheric neutrino experiments. This scenario
is accompanied by the small mixing angle MSW solution involving 
$\nu_e-\nu_s$.
We now consider how an alternate description is possible in 
which $\nu_{\mu }-\nu_s$
oscillations resolve the atmospheric anomaly, while 
$\nu_e-\nu_{\tau }$ vacuum
oscillations resolve the solar neutrino puzzle. It turns out that this
`bi-maximal' mixing scenario permits the existence of
neutrino 
hot dark matter, with 
contribution to the cosmological density parameter of 
$0.2-0.25$. 
Models of
structure formation with both cold and hot dark matter \cite{cdmat} 
are in good agreement with the observations.
 
We introduce the MSSM singlet states ${\cal N}_{1,3}$ and $\nu_s$, 
such that

\begin{equation}
Q_{{\cal N}_1}=-(p'+r')~,~~~~~ Q_{{\cal N}_3}=p'~, 
\label{QN13} 
\end{equation}
($0< p' \leq n+k$,~ $r'\geq 0$).
The relevant superpotential couplings involving the ${\cal N}$ states are

\begin{equation}
\begin{array}{cc}
 & {\begin{array}{cc}
{\cal N}_1~~~~~&\,\,{\cal N}_3~~~~~~
\end{array}}\\ \vspace{2mm}
\begin{array}{c}
l_1\\ l_2 \\l_3  
 
\end{array}\!\!\!\!\! &{\left(\begin{array}{ccc}
\,\, \epsilon^{n+k+p'+r'} &
\,\,  \epsilon^{n+k-p'}
\\ 
\,\, \epsilon^{n+p'+r'}~~~ &\,\,0~~~~~     
\\
\,\, \epsilon^{p'+r'}~~~ &\,\,0~~~~~     
\end{array}\right)\kappa' h_u }~, 
\end{array}  \!\!~~~~~
\begin{array}{cc}
 & {\begin{array}{cc} 
{\cal N}_1~~~&\,\,
~~{\cal N}_3~~~ 
\end{array}}\\ \vspace{2mm}
\begin{array}{c}
{\cal N}_1 \\ {\cal N}_3
 
\end{array}\!\!\!\!\! &{\left(\begin{array}{ccc}
\,\, \epsilon^{2(p'+r')}
 &\,\,\epsilon^{r'}
\\ 
\,\, \epsilon^{r'}~~~~~~
&\,\,0
\end{array}\right)M'~,
} 
\end{array}~~~
\label{nuNNN1}
\end{equation}
The sterile state $\nu_s$ is required to have large mixing with 
$\nu_{\mu }$. 
This means that
its ${\cal U}(1)$ charge must be such that it avoids forming a 
'heavy' state with
$\nu_{\tau }$. This is easily achieved if $1\leq Q_{\nu_s}\leq  n$,
in which case the mass term for $\nu_s$ will also be forbidden. 
Consider the
superpotential couplings

\begin{equation}
W_{\nu_{e, \mu }\nu_s }=\Gamma '(l_2+
\left(\frac{X}{M_P}\right)^kl_1)h_u.
\label{numus}
\end{equation}
Taking into account the couplings (\ref{nuNNN1}), (\ref{numus}) and 
integrating out the heavy ${\cal N}_{1,3}$ states, the mass 
matrix for the active and sterile neutrinos takes the form:

\begin{equation}
\begin{array}{cccc}
 & {\begin{array}{cccc} 
\hspace{-5mm}~~~~~\nu_e~ & \,\,~~~~~~\nu_{\mu }~~~  & \,\,~\nu_{\tau }~~
& \,\,\nu_s
\end{array}}\\ \vspace{2mm}
m_{\nu }= \begin{array}{c}
\nu_e \\ \nu_{\mu } \\ \nu_{\tau } \\ \nu_s
 \end{array}\!\!\!\!\! &{\left(\begin{array}{cccc}
\,\,m'\epsilon^{n+k}  &\,\,~~m'\epsilon^n &
\,\,~~m' &~~m\epsilon^k
\\ 
\,\,m'\epsilon^n~~  &\,\,~~0  & \,\,~~0&m
 \\
\,\, m'~ &\,\,~~ 0 &\,\,~~0 & 0
\\
\,\, m\epsilon^k &\,\,~~ m &\,\,~~0 &m_s
\end{array}\right) }~, 
\end{array}  \!\!  ~~~~~
\label{nus}
\end{equation}
where $m'=\frac{\kappa'^2\langle h_u\rangle^2 }{M'}\epsilon^{n+k}$, 
and we have defined
$m\equiv \Gamma '\langle h_u\rangle $.
$m_s$ is the mass of the sterile neutrino. Farther
we will assume that $m_s\ll m$.
The required suppressions of $\Gamma '$ and $m_s$ can be achieved 
with the help of ${\cal U}(1)$ symmetry (\cite{chun,61,422}).

From (\ref{nus}) we see that $\nu_e-\nu_{\tau }$
and $\nu_{\mu }-\nu_s$ form massive states,  
\begin{equation}
m_{\nu_1 }\simeq m_{\nu_3 }\simeq m'~,~~~~
m_{\nu_2}\simeq m_{\nu_s }\simeq m~,
\label{deg132s}
\end{equation}
and for the oscillation parameters we obtain:

\begin{equation}
\Delta m_{e\tau  }^2\sim 2m'^2\epsilon^{n+k}~,~~~~
\sin^2 2\theta_{e\tau }= 1-{\cal O}(\epsilon^{2(n+k)}),
\label{shift13}
\end{equation}

\begin{equation}
\Delta m^2_{\mu s  }\simeq 2m(m'\epsilon^{n+k}+m_s)~,~~~~
\sin^2 2\theta_{\mu s}= 1-{\cal O}(\frac{m_s^2}{m^2}).
\label{shift2s}
\end{equation}       


To obtain $\Delta m_{e\tau }^2\simeq 10^{-10}~{\rm eV}^2$ in
(\ref{shift13}), for $1\stackrel{_<}{_-}n+k\stackrel{_<}{_-}5$,
the $m'$ should be taken in the range 
$\sim 2\cdot 10^{-5}-4\cdot 10^{-4}$~eV respectively.
The first term in (\ref{shift2s}) does not give significant contribution 
to the  $\Delta m^2_{\mu s }$ (for allowed   
values of $m'$
and $n+k$). The required splitting can 
occur due to $m_s$. For 
$m_s\sim 5\cdot 10^{-4}$~eV, $m$ can be $\sim 3$~eV, corresponding
to neutrino hot dark matter
$\sim 15\%$ of the critical energy density.
For larger $m_s$ 
(say $\sim 2\cdot 10^{-2}$~eV), $m$ should be taken less then 
$\sim 1$~eV, which implies a non-significant contribution 
to dark matter.


 In summary, the existence of   
neutrino hot dark matter is predicted in 
models in which $\nu_{\mu }-\nu_s$ oscillations with large 
mixing explain 
the atmospheric neutrino anomaly, while the large angle 
$\nu_e-\nu_{\tau }$ vacuum
oscillations resolve the solar neutrino puzzle.

In obtaining maximal mixing between `active'
neutrinos the forms of matrices
in (\ref{nu2}), (\ref{nus}) is crucial, and for the mechanism 
to work, one has to make sure that the `zero' entries in these 
matrices will not be (radically) changed. The zero entries 
were guaranteed by the
${\cal U}(1)$ symmetry (see couplings in (\ref{mats}), (\ref{nuNNN1})), 
and in $\nu $MSSM 
there is no additional source which could
change the picture. However, in models such as flipped $SU(5)$,
$SU(4)_c\times SU(2)_L\times SU(2)_R$, or $SO(10)$ ,
in addition to the MSSM `matter' content, we also have
the MSSM singlet states $\nu^c_i$ with `Dirac' couplings 
to the left handed neutrinos. Through the seesaw mechanism, they
give rise to an additional contribution $m_{\nu }'$ to the
neutrino mass matrix. The magnitudes of the appropriate 
entries of $m_{\nu }'$ depend on the specifics of the model,
but for the scenario to work, these must be
$\stackrel {_<}{_\sim } m\epsilon^n$ 
(see (\ref{nu2})). 
One way to achieve this is to introduce 
additional `matter' singlet ($N$) states.
If $N$ and $\nu^c$ superfields form large massive states  
$M_0\nu_cN$, and the terms $lNh_u$ and $N^2$ either do not
exist or are strongly suppressed (this could be arranged by the 
${\cal U}(1)$ symmetry), it is easy to verify that after 
integrating out of the heaviest $\nu^c-N$ states, the neutrino 
masses will not receive significant contributions from the
`physics of $\nu^c$ sector', and the large $\nu_{\mu }-\nu_{\tau }$
($\nu_e-\nu_{\tau }$) mixings will be realized. 
We will demonstrate this shortly with an explicit example based
on flipped $SU(5)$.

Before discussing this case, however, let us mention
another possible scenario in which $\nu_e-\nu_{\tau }$ small
angle MSW oscillations  resolve the solar neutrino puzzle. 
Since the $\nu_{\tau }$ state must be 
light it should not mix strongly with $\nu_{\mu }$.
Since in this case we do not deal with large mixings between the active
neutrino states we do not need to apply mechanism presented (involving
${\cal N}$ states) above.
It turns out that this case predicts the heaviest neutrinos
to be $\stackrel{_<}{_\sim }$~eV. 
Within the framework of $\nu$MSSM and 
some of its extensions, the mass matrix of the active neutrinos
is expected to be as given by (\ref{mactive}),
with $\nu_e-\nu_{\tau }$ mixing $\sim \epsilon^{n+k}$. 
For small angle MSW solution,
a good choice is $n+k=2$, and 
$m'(=m_{\tau })\sim (10^{-3}-5\cdot 10^{-3})$~eV, which means that  
we need a sterile state
$\nu_s$ to explain the atmospheric anomaly. 
To avoid  large $\nu_{\tau}-\nu_s$ mixing, $n$ must differ from zero.
With the help of ${\cal U}(1)$ symmetry, 
$\nu_{\mu}$ will form a `degenerate' massive state with $\nu_s$  
($m_{\nu_2 }\simeq m_{\nu_s }$). Having Couplings
$\Gamma''(\nu_{\mu }+\epsilon^k\nu_e)\nu_sh_u\equiv 
m(\nu_{\mu}+\epsilon^k\nu_e)\nu_s$, the whole neutrino mass matrix will
have the form:

\begin{equation}
\begin{array}{cccc}
 & {\begin{array}{cccc} 
\hspace{-5mm}~~~~~\nu_e~ & \,\,~~~~~~~~~~\nu_{\mu }~~~  & 
\,\,~~~~~\nu_{\tau }~~& \,\,~~~\nu_s
\end{array}}\\ \vspace{2mm}
m_{\nu }= \begin{array}{c}
\nu_e \\ \nu_{\mu } \\ \nu_{\tau } \\ \nu_s
 \end{array}\!\!\!\!\! &{\left(\begin{array}{cccc}
\,\,m'\epsilon^{2(n+k)}  &\,\,~~m'\epsilon^{2n+k} &
\,\,~~m'\epsilon^{n+k} &~~m\epsilon^k
\\ 
\,\,m'\epsilon^{2n+k}~~  &\,\,~~m'\epsilon^{2n}  & \,\,~~m'\epsilon^n&m
 \\
\,\, m'\epsilon^{n+k}~ &\,\,~~m'\epsilon^n &\,\,~~m' & 0
\\
\,\, m\epsilon^k &\,\,~~ m &\,\,~~0 &m_s 
\end{array}\right) }~, 
\end{array}  \!\!  ~~~~~
\label{mactst}
\end{equation}
where 
we will still assume that $m_s\ll m$.

For the oscillation parameters we will have:


\begin{equation}
\Delta m_{\mu s}^2\simeq m(3m'\epsilon^{2n}+2m_s)~,~~~
\sin^2 2\theta_{\mu s} =1-{\cal O}(\epsilon^{2n})-
{\cal O}(\frac{m_s^2}{m^2})~,
\label{dif}
\end{equation}

\begin{equation}
\Delta m_{e\tau }^2\sim m'^2~,~~~
\sin^2 2\theta_{e\tau } \sim 4\epsilon^{2(n+k)}~~~(n+k=2)~.
\label{difsol}
\end{equation}
Two cases which give different implications should be considered:
({\it i}) For $m_s\sim 5\cdot 10^{-4}$~eV in order to get
$\Delta m_{\mu s}^2\sim 3\cdot 10^{-3}~{\rm eV}^2$ (for either 
$n=1$ or $2$) the value for $m$ should be taken 
(according to (\ref{dif})) $m\sim 3$~eV;
({\it ii }) For an increased value of $m_s$, 
say $\sim 1.5\cdot 10^{-3}$~eV, (for either $n=1$ or $2$)
$m$ should be taken $m\sim 1$~eV, which is 
already not enough to have any appreciable neutrino hot dark matter. 
As we see the case ({\it i}) gives the possibility of existence of the
neutrino hot dark matter also in this scenario.





%
%

To demonstrate how the scenarios discussed above can be realized 
in realistic
models beyond $\nu$MSSM and, in particular, how the mechanisms work
in practice, let us
consider the flipped $SU(5)$ model \cite{flip}. We picked this 
model especially since the ${\cal U}(1)$ charges of the `matter' 
superfields 
are rather precisely determined.
The matter content of the model is
\footnote{We assume the existence of $Z_2$ `matter' parity which
distinguishes the matter from the higgs supermultiplets.}:

\begin{equation}
10_i=(q,~d^c,~\nu^c )_i~,~~~~
\bar 5_i= (u^c,~l)_i~,~~~~
1_i= e^c_i~,
\label{repflip1}
\end{equation}
where $i=1, 2, 3$ denotes the family index.
Their `gauge' $U(1)$ charges are $1$, $-3$ and $5$ respectively.
Similarly  the `higgs' multiplets are :

\begin{equation}
\phi (5)=(h_d,~\bar D^c)~,~~~
\overline{\phi } (\bar 5)=(h_u,~ D^c)~,~~~
H (10)~,~~~\bar H(\overline{10})~,
\label{repflip2}
\end{equation}
with $U(1)$ charges $-2$, $2$, $1$ and $-1$ respectively.
The VEVs of the scalar components of $H, \bar H$ superfields break 
$SU(5)\times U(1)$ to $SU(3)_c\times SU(2)_L\times U(1)_Y$. 
The well known doublet-triplet splitting 
problem is solved through the missing partner mechanism \cite{flip}.

The superpotential couplings which generate the quark, 
`Dirac' and charged 
lepton masses are:

\begin{equation}
W_Y={\cal A}_{ij}10_i10_j\phi +
{\cal B}_{ij}10_i\bar 5_j\overline{\phi } +
{\cal C}_{ij}\bar 5_i1_j\phi~,
\label{yuk}
\end{equation}
where $i, j$ are family indices.
The matrices
${\cal A}$, ${\cal B}$, ${\cal C}$ have entries that depend on 
powers of $\epsilon $. 
It is easy to see that the desired hierarchy (\ref{lambdaud})
for the down quarks is achieved if
the ${\cal U}(1)$ charges 
of $10_i$  are related as follows:  
                                     
\begin{equation}
Q_{10_3}-Q_{10_2}=Q_{10_2}-Q_{10_1}=1~.
\label{Q10}
\end{equation}                       
From (\ref{Q10}) and the known hierarchies of the up quark 
Yukawa couplings (\ref{lambdaud}), we also have:
                                                     
\begin{equation}                                     
Q_{\bar 5_3}-Q_{\bar 5_2}=Q_{\bar 5_2}-Q_{\bar 5_1}=2~.
\label{Q5}
\end{equation}                       
Similarly, from (\ref{lamdae}) and (\ref{Q5}) we find:

\begin{equation}                                     
Q_{1_3}=Q_{1_2}~,~~~~~Q_{1_2}-Q_{1_1}=1~.
\label{Q1}
\end{equation}                       
Given (\ref{Q10}), (\ref{Q5}) and (\ref{Q1}), we can determine
the structure 
of the Yukawa matrices ${\cal A}, {\cal B}, {\cal C}$:
                                                     
$$
{\cal A}_{ij}\sim \epsilon^{a+6-i-j}~,~~~~~~~
{\cal B}_{ij}\sim \epsilon^{9-i-2j}~,
$$
\vspace{0.2cm}
\begin{equation}
\begin{array}{ccc}
{\cal C}_{ij}\sim~~ \\
\end{array}
\hspace{-6mm}\left(
\begin{array}{ccc}
\epsilon^5& \epsilon^4 &
 \epsilon^4 \\
\epsilon^3&\epsilon^2 &
\epsilon^2  \\
\epsilon & 1 & 1  \end{array}
\right)\epsilon^a~,~~~~
\label {ABC}
\end{equation}
where $0\leq a\leq 2$ determines the value of the MSSM parameter 
$\tan \beta $($\sim \frac{m_t}{m_b}\epsilon^a$).

Since the left handed leptons are contained in the $\bar 5$-plets,
from (\ref{Q5}) and (\ref{lep3}), we see that we are dealing with the
case $(n,~k)=(2, 2)$. Therefore, without some suitable extension 
of this minimal scheme, we expect the $\nu_{\mu }-\nu_{\tau }$
mixing angle $\theta_{\mu \tau }\sim \epsilon^2$, in strong
disagreement 
with the data.
The choice $(n,~k)=(2, 2)$ tells us that we should suitably enlarge the
model, so as to accommodate large 
angle $\nu_{\mu }-\nu_{\tau }$ oscillations for the atmospheric anomaly,
while the solar neutrino 
puzzle can be resolved through the small mixing angle (MSW) 
oscillations of $\nu_e$ into a sterile state $\nu_s$ 
(the other two scenarios, it turns out, 
either give  $m_{\nu_{\mu }}\sim 1$~keV or
$\theta_{e\tau}\sim \epsilon^4$, both of which are unacceptable). 


To implement this scenario, we will take the 
horizontal ${\cal U}(1)$ symmetry to be an 
${\cal R}$ symmetry.
This will be crucial for the natural generation of appropriate mass 
scales in the neutrino sector.
We introduce the singlet states ${\cal N}_{2}$, ${\cal N}_{3}$ and 
$\nu_s$. Let us prescribe the following  
${\cal R}$ charges to the 
various superfields:

$$
R_{10_i}=(R-R_{\phi })/2-(4-i)R_X~,~~~
R_{\bar 5_i}=(R+R_{\phi })/2-R_{\overline{\phi }}-
(5-2j)R_X~,
$$
$$
R_{1_1}=(R-3R_{\phi })/2+R_{\overline{\phi }}-4R_X~,~~~
R_{1_2}=R_{1_3}=R_{1_1}+R_X
$$
\begin{equation}
R_{{\cal N}_2}=(R-R_{\phi })/2-R_H-R_X~,~~~
R_{{\cal N}_3}=R_{{\cal N}_2}+2R_X~,~~~~R_{\bar H}=R_{\phi }/2~,
\label{QR}
\end{equation}
where $R_{\phi }, R_{\overline{\phi }}, R_H$ and $R_X$ remain undetermined.
This choice corresponds to $a=2$ in (\ref{ABC}),
which means that the MSSM parameter $\tan \beta $ is not too
far from unity. 

The superpotential couplings involving ${\cal N}_2$, ${\cal N}_3$ 
fields are

\begin{equation}
\begin{array}{cc}
 & {\begin{array}{cc}
{\cal N}_2&\,\,{\cal N}_3~~   
\end{array}}\\ \vspace{2mm}
\begin{array}{c}
\bar 5_2\\ \bar 5_3 \\
\end{array}\!\!\!\!\! &{\left(\begin{array}{ccc}
\,\, \epsilon^2~~ &
\,\, 1
\\
\,\, 1~~ &\,\,0
\end{array}\right)\frac{\overline{\phi }H}{M_P} }
\end{array}  \!\!~,~~~~
\begin{array}{cc}
 & {\begin{array}{cc}
{\cal N}_2 &\,\,
{\cal N}_3~~~~~~~~~~~~~                         
\end{array}}\\ \vspace{2mm}
\begin{array}{c}
{\cal N}_2 \\ {\cal N}_3

\end{array}\!\!\!\!\! &{\left(\begin{array}{ccc}
\,\, \epsilon^2~~
 &\,\,1
\\
\,\,  1~~
&\,\,0
\end{array}\right)\left(\frac{\bar HH}{M_P^2}\right)^2M_P}
\end{array}~.~~
\label{Nflip}
\end{equation}
We see from (\ref{Nflip}) that the coefficient 
$\kappa \sim \epsilon_G$($= M_{GUT}/M_P$), the mass 
scale $M\sim M_P\epsilon_G^4$ 
(see (\ref{mats})), and  
$m\sim \kappa^2\langle h_u\rangle^2/M\sim 
\langle h_u\rangle^2/(M_P\epsilon_G^2)\simeq 0.13$~eV, which
is precisely what is needed for the $n=2$ case 
(see discussion after (\ref{atm})).
Here $p=1,~r=0$ and 
$n\to n+1,~k\to k+1$ (what matters are the relative values
of the ${\cal U}(1)$ charges).
After integrating out the ${\cal N}$ states, the 
$\nu_{\mu }-\nu_{\tau }$
mass matrix will have precisely the desired form (\ref{nu2}).
We therefore should expect
$\nu_{\mu }-\nu_{\tau }$ oscillations with the 
parameters given by (\ref{atm}). 


The allowed couplings 
$\left(\frac{X}{M}\right)^{8-i-j}10_i10_j\bar H^2/M_P$ generate 
masses for $\nu^c$ states 
which, through the seesaw mechanism, yield contribution to the 
(2,2) element of matrix (\ref{nu2})
$\sim \langle h_u\rangle^2/(M_P\epsilon_G^2\epsilon^2) 
\simeq 3$~eV. This would spoil the picture of maximal mixing in the
$\nu_{\mu }-\nu_{\tau }$ sector.
For things to work out, we introduce additional 
singlet superfields $N_i$ (one per generation), with ${\cal R}$ charges
$R_{N_i}=R/2+(4-i)R_X$.
With the couplings $\left(\frac{X}{M_P}\right)^{i-j}N_i10_j\bar H$ 
permitted, but 
the couplings $N_iN_j$ and $N_i\bar 5_j\cdot \overline{\phi }H/M_P$
forbidden by the ${\cal R}$-symmetry, integration of the 
heaviest $N-\nu^c$ 
states will not yield any contribution to the light neutrino 
masses. We have also checked that after integration of $N-\nu^c$ states,
the forms of couplings (\ref{Nflip}) are not changed. In other words, 
maximal $\nu_{\mu }-\nu_{\tau }$ mixing through our mechanism
can be realized in flipped $SU(5)$.

The resolution of the solar neutrino puzzle requires a sterile state
$\nu_s$ with ${\cal R}$ charge  given by 
$R_{\nu_s}=(R-R_{\phi })/2-R_H-17R_X$. The relevant couplings are:

\begin{equation}
\left(\frac{X}{M_P} \right)^{20}
\frac{H}{M_P}\bar 5_1\nu_s\overline{\phi }+
\left(\frac{\bar HH}{M_P^2} \right)^2
\left(\frac{X}{M_P} \right)^{34}M_P\nu_s^2~,
\label{stflip}
\end{equation}
from which, the corresponding mass scales are
\begin{equation}
m_{\nu_e \nu_s}\sim
(2\cdot 10^{-5}-10^{-4})~{\rm eV}~,~~
m_{\nu_s }\sim (4\cdot 10^{-5}-10^{-3})~{\rm eV}.
\label{oscilflip}
\end{equation} 
Comparison with expression (\ref{nust}) shows that the solar neutrino
puzzle is resolved via the small mixing angle MSW solution.

In conclusion, we have presented a rather general scheme 
for realizing maximal (and even bi-maximal) mixing in realistic 
models such as $\nu$MSSM and supersymmetric GUTs. Except in 
special cases, we are led to introduce a sterile neutrino so as 
to resolve the solar and atmospheric neutrino
puzzles \cite{lsnd}. A ${\cal U}(1)$ flavor symmetry, especially 
if it happens 
to be a ${\cal R}$-symmetry, can nicely protect the sterile neutrino 
from becoming heavy. This symmetry, in addition, helps provide a 
consistent framework for understanding the charged fermion mass 
hierarchies, as well as the magnitudes of the CKM matrix elements. 
Neutrino hot dark matter can arise in models which
exhibit either 
bi-maximal  neutrino mixing or maximal $\nu_{\mu }-\nu_s$ and small
angle $\nu_e-\nu_{\tau }$ MSW oscillations.

%
%

\end{document}